\documentclass[preprint]{aastex}
\input{psfig}
\begin{document}
\def\etal{{\it et al.\/}}
\def\cf{{\it cf.\/}}
\def\ie{{\it i.e.\/}}
\def\eg{{\it e.g.\/}}

\title{On particle acceleration around shocks. I}
\author{{\bf Mario Vietri}}
\affil{Scuola Normale Superiore, Pisa}
{}
\begin{abstract}
We derive a relativistically covariant (although not manifestly so)
equation for the distribution function of
particles accelerated at shocks, which applies also to extremely
relativistic shocks, and arbitrarily anisotropic particle distributions. The
theory is formulated for arbitrary pitch angle scattering, and reduces to the
well--known case for small angle scatterings via a Fokker--Planck approximation.
The boundary conditions for the problem are completely reformulated introducing
a physically motivated Green's function; the new formulation allows derivation
of the
particle spectrum both close and far away from the injection energy in an exact
way, while it can be shown to reduce to a power--law at large particle energies.
The particle spectral index is also recovered in a novel way. Contact is made
with the Newtonian treatment.
\end{abstract}


\section{Introduction}

The theory of particle acceleration at shocks is currently unsatisfactory.
We cannot follow this process from first principles, \ie, the development
of a collisionless shock, with the ensuing injection of non-thermal particles,
nor do we understand the later shock evolution, which should include the
interaction between the population of high--energy, non--thermal 
particles, and the shock structure. Even in the test--particle limit, to
which this paper is totally confined, there are at least two major problems. 
First, the properties of the scattering agents are very
poorly known; this problem will not be dealt with in this paper. Second,
the general foundations of the theory have not been laid down. In particular,
there is no general equation for the particle distribution function (DF) for
arbitrary (\ie, even relativistic) shock speed, and no general treatment is
available when the DF is anisotropic. Most results available to us either have
been derived in the vanishing speed (thus fully isotropic) limit (Bell 1978,
Blandford and Ostriker 1978), or descend from numerical (Bednarz and Ostrowski,
1998) or semi-numerical (Kirk and Schneider 1987) approaches. This is true
even in the idealized situation of a plane shock expanding forever in a uniform
medium,with particle scattering due only to pitch--angle scattering. It is the
aim of this paper to derive a sufficiently general equation describing in a
relativistically covariant way this process, to elucidate the conditions
leading to the determination of the particle spectrum, and to apply these
results to the extremely relativistic limit, where the DF anisotropy is
expected to be maximal. In passing, a number of old results will be recovered:
it will be shown that the probability of reaching infinity is independent
of the particle momentum, that the spectrum is the superposition of many
bumps, each corresponding to a set of particles which have crossed the shock
$0,1,2,...,N,..$ times, and that, for very large particle energy,
the superposition of these bumps  spectrum leads to a power law in the 
particle impulse: all these results will be shown to
hold for all shock speeds, Newtonian, relativistic or intermediate ones.
Also, the explicit dependence upon the injection spectrum will be 
presented. 
In a future paper, this formalism will be applied to the extremely relativistic
limit, including the effects of small or large angle scatterings, and a 
mean magnetic field. 

The plan of this paper is as follows. In Section 2, we derive the new
equation for the particle DF, in an explicitly relativistic covariant
way. In Section 3, we reformulate the boundary conditions to which this
problem is subject, in a physically motivated way, finding eventually 
the dependence of the distribution function upon the injection spectrum.
In Section 4, we concentrate on the distribution function at energies
large compared with those of injection; here we show that it must 
be a power law, showing what fixes the power law index analytically,
and deriving the relativistic generalization of Bell's law. We also
make contact with the Newtonian limit. The important probability 
distributions $P_d$ and $P_u$, whose existence is merely postulated in
the previous sections, are explictly written out in terms of eigenfunctions
of the angular part of the scattering equation in Section 5.
In Section 6, we summarize our results.

\section{An equation for the distribution function}

We consider a shock propagating at arbitrary speed in an homogeneous fluid.
We place ourselves in the shock reference frame, and choose coordinates
such that $z = 0$ identifies the shock position. The upstream fluid is
located at $z<0$, so that the fluid speed is always $> 0$, both
upstream and downstream. Suitable jump conditions at the shock are
assumed (Landau and Lifshitz 1987), but they will not be necessary
in this section.

We assume now that we can break down the effects of the
magnetic field into two parts, as is customarily done. The first part is
due to a small--coherence length magnetic field, possibly self--excited
by the particles; this provides for an effective scattering,
which is included in the collisional term. Another part, however,
is due to a long--coherence length field, which provides instead for a
smooth deflection of particles in phase space; this effect is included
in the convective term (see below). We are thus using a dichotomic
description of the
magnetic field; most of the results to be described here, which are
wholly unavoidable within this description, ultimately depend on the
correctness of this assumption on the magnetic field.

The distribution function is of course subject to the collisionless Vlasov
equation, supplemented with suitable collisional and injection terms. In
somewhat abstract form, we have
\begin{equation}
\label{abstract}
\frac{d f}{d s} = \left(\frac{\partial f}{\partial s} \right)_{coll} + \phi_{inj}\;.
\end{equation}
The right hand side is the collisional term, with collisions per unit proper
time, plus the rate of particles injected at the shock.

The convective term is
\begin{equation}
\label{convective}
\frac{d f}{d s} \equiv \frac{d x^\nu}{d s}\frac{\partial f}{\partial x^\nu} +
\frac{d p^\nu}{d s} \frac{\partial f}{\partial p^\nu}\;.
\end{equation}
In the above equation, the term containing the derivative with respect to
the energy is not usually explicitly written out, imposing instead
{\it ab initio} that particles be on their mass shell; here we do this
instead (showing later on that it makes no difference) in order to
emphasize the covariant character of the equation. In fact,
both the first and the second term on the right hand side are invariant for
Poincar\'e group transformations. We make use of this by computing each term
in different frames of reference, and with different orientations for the
coordinate axes. In the first term, we place ourselves in the shock frame and
orient the $z$ axis along the shock normal, so that we can make
use of $\partial/\partial t = \partial/\partial x = \partial/\partial y = 0$,
because we assume the shock to be in steady state, and to have planar symmetry.
We also write the $z$ component of the particle four--speed in terms of
quantities in the fluid frame,
\begin{equation}
\frac{d z}{d s}= \gamma_p \gamma (u+\mu)\;,
\end{equation}
where $\mu$ is the cosine of the angle which the particle speed makes with the
shock normal, $z$, in the fluid frame, $u$ and $\gamma$ are the shock
speed (in units of $c$) and Lorentz factor speed with
respect to the fluid, and $\gamma_p$ is the particle Lorentz
factor also with respect to the fluid. All particles are here assumed
relativistic in any reference frame (so that their three-speeds are always
$\approx c$), even those which are just injected. Strictly speaking, this
is accurate only for relativistic shocks. The generalization to include 
Newtonian particles, though easy in principle to do, introduces some
cumbersome formulae (with radicals) which complicate this work uselessly:
it will not be considered any further. 

The second term on the right hand side of Eq. \ref{convective} will be instead
computed in the fluid frame, so that we may neglect any motional
electric field. In this case, the particle energy is conserved, and the term
including the derivative $\partial f/\partial E$ disappears. The remaining
components can be written as (Berezinsky {\it et al.}, 1990, Ch.IX, Eq. 9.16)
\begin{equation}
\label{simpl2}
\frac{d p^\nu}{d s} \frac{\partial f}{\partial p^\nu} =
\frac{d t}{d s} \frac{d p^\nu}{d t} \frac{\partial f}{\partial p^\nu} =
- \gamma_p \omega \frac{\partial f}
{\partial \phi}
\end{equation}
where $\omega = e B/ E$ is the particle Larmor frequency in the fluid
frame, and $\phi$ is the longitudinal angle around the direction of the
magnetic field, in {\bf momentum} space.

To estimate the collisional term, we first place ourselves in the fluid
frame, and then remark that, in full generality, such term has the form
\begin{equation}
\label{coll}
\left(\frac{\partial f}{\partial t}\right)_{coll} =
- d(\mu) f + \int w(\mu, \mu') f(\mu') d\!\mu'\;.
\end{equation}
Here, we have made use of the assumption of linearity: we assume, that is, that the
magnetic field turbulence responsible for the scattering is independent of the
particle density. We consider particle scattering when there may be a correlation
between the directions before and after the scattering event, so that the 
scattering coefficient $w(\mu,\mu')$ depends on both. Also, we are not assuming 
the diffusion approximation: the above equation applies to arbitrarily few, and
arbitrarily large--angle, scattering laws. It would reduce to the
well--known Ginzburg--Syrovatskii (1964) form in the diffusive approximation. 
The first term on the right hand side represents scattering
of particles away from their direction of motion, while the second term
represents the total contribution to the distribution function in the direction
of motion under consideration, by scattering from all other directions.
Clearly, $d$ and $w$ are related through
\begin{equation}
\label{constraint}
d(\mu) = \int_{-1}^{+1} w(\mu', \mu) d\!\mu'
\end{equation}
which simply expresses probability conservation. $w$ (and thus also $d$)
may depend upon $p$. Often, one makes the simplifying assumption that the
$p$--dependence in $w$ (and thus in $d$) factors out; we shall comment
later on why this may come in handy, but will not assume this.
We also remark that $w,d > 0$.

Since we need $\partial f/\partial s$, while above (Eq. \ref{coll}) we have
evaluated the collisional term in the fluid frame, we correct it by means of
$d t/d s = \gamma_p$. Also, we wish to include the injection term, $\phi_{inj}$.
We assume that injection takes place at the shock only, but we leave the
energy and direction dependence of the term otherwise arbitrary: $\phi_{inj}
= G_{inj}(p,\mu) \delta(z)$ where $\delta(x)$ is Dirac's delta.

Putting together all of the above, we find
\begin{equation}
\label{refer}
\gamma (u + \mu) \frac{\partial f}{\partial z} = \left( - d(\mu) f +
\int w(\mu, \mu') f(\mu') d\!\mu'\right) + \omega \frac{\partial f}
{\partial \phi} + G_{inj}(p,\mu) \delta(z)\;,
\end{equation}
which is the equation we were searching for. It is worth remarking that all
terms in the above equation are computed in the fluid frame, except for the
space dependence, and its attaining derivative.Also, the above equation is
valid both upstream and downstream, but now care must be taken because
the quantities $p,\mu,\mu',\omega,... $ are all evaluated in the local fluid
frame, and thus quantities bearing the same name in the two distinct reference
frames are not identical: they are related instead by
a Lorentz transformation, which we give here for future reference. Using
subscript $u$ and $d$ to denote quantities in upstream and downstream frame
respectively, he have
\begin{equation}
\label{transf}
p_u = p_d \gamma_r (1-u_r \mu_d) \;;\;
\mu_u = \frac{\mu_d - u_r}{1-u_r \mu_d}\;,
\end{equation}
where $u_r$ and $\gamma_r$ are the {\bf modulus} of the relative speed
between the upstream and downstream sections, and its associated Lorentz factor.
{}

Whenever the scattering in each individual event is small, it is
possible to apply to the above equation the very same treatment which
leads to a Fokker--Planck equation. In this case one obtains (Kirk and Schneider
1987)
\begin{equation}
\label{refer2}
\gamma (u + \mu) \frac{\partial f}{\partial z} =
\frac{\partial}{\partial \mu}\left(D(\mu,p) (1-\mu^2)
\frac{\partial f}{\partial\mu}\right)
+ \omega \frac{\partial f}
{\partial \phi} + G_{inj}(p,\mu) \delta(z)\;,
\end{equation}
where the effective diffusion coefficient is related to the r.m.s. angle
of diffusion.
It is worth remarking that it may occur that the above equation holds for the
downstream section, but it is not justified for the upstream section, where
the full Eq. \ref{refer} has to be employed. The reason is that the
diffusive approximation on which the above equation is based requires each
scattering event to lead to a r.m.s. deflection ${\cal O} (1/\gamma)$, where
$\gamma$ is the fluid Lorentz factor with respect to the shock. When the shock
becomes extremely relativistic, thus, the correctness of the diffusive
approximation breaks down: it follows that one may use Eq. \ref{refer2} for the
downstream section, but is forced to use Eq. \ref{refer} for the upstream one.

In the following, we drop the term due to the long--coherence length magnetic
field, which will be considered in Paper II. 

\section{Obtaining the particle spectrum}

We begin by remarking that in this problem there is a net mean flux
of particles across any surface; if the surface is steady in the
shock frame, then the flux is also time--independent. Since the
particles' speed is always assumed $\approx c$ ($\approx 1$ in our
units), the infinitesimal flux across a unit shock area, per unit
time, is given by
\begin{equation}
\label{fluxdef}
d\!J = \mu_s p_s^2 f d\!\mu_s d\!p_s =
\gamma_u (u+\mu_u) p_u^2 f d\!\mu_u d\!p_u =
\gamma_d (u_d+\mu_d) p_d^2 f d\!\mu_d d\!p_d
\end{equation}
depending upon whether we choose to express $f$ in terms of
variables in the shock, upstream, or downstream frames. The representation
in terms of downstream variables is especially useful, because in the
downstream section the downstream momentum $p_d$ is conserved, so that a
net flux across the shock of particles of impulse $p_d$ can be due only
to the fact that some fraction of these particles is advected to downstrean
infinity, with probability given by Eq. \ref{prob}.

From now on, in this section, we shall use coordinates in the downstream frame
exclusively, which we thus indicate, for ease of notation, without subscripts.

The mathematical treatment of this problem must be {\bf physically} motivated:
we envision a system where particles are injected at low energies at the shock,
from which they cannot reach upstream infinity, because they would have to swim
upstream, but from which they do reach downstream infinity. In order to study
how particles generated at the shock reach downstream infinity, we first
concentrate on the downstream part of the problem.

Consider now a downstream section of finite length $L$, beginning at the shock;
then place two observers at either end, and have them inject particles into the
downstream section. Then they may collect outgoing particles at either end. The
problem is obviously physically well defined. The speed of a particle (in 
downstream frame coordinates) with respect to the shock is
\begin{equation}{}
v_s = \frac{u_d+\mu}{1+u_d\mu}\;.
\end{equation}
Thus, the observer
located at the shock injects particles for $\mu > -u_d$ (here $u_d$ is the
downstream fluid speed with respect to the shock), and collects the
outgoing ones (\ie, those with $\mu < -u_d$). So reasonable boundary conditions
at the shock can be posed for $\mu > -u_d$ only. The observer at the other end
of the downstream section disappears as $L \rightarrow \infty$, so that there
can be no entering flux there; we can only impose the regularity condition that
$f$ does not grow to infinity, nor goes to zero.

Similarly, for the upstream section we see that we can impose boundary
conditions for $\mu< -u_d$ at the shock; at upstream infinity, the only
reasonable boundary condition is that $f\rightarrow0$, because we cannot expect
our particles to swim against the fluid advection all the way to upstream
infinity.

We now see that the boundary conditions for the upstream problem are provided
by the outgoing particles of the downstream section, and {\it vice versa} the
outgoing particles of the upstream section provide the boundary conditions of
the downstream section, thusly setting up an obvious (fixed point) problem. In
order to make this explicit, let us introduce the conditional probability 
$P_d(\mu_{in},\mu_{out}) d\!\mu_{out}$ that a particle, given that it crossed
the shock toward downstream along a direction $\mu_{in}$, will recross it along 
the direction $\mu_{out}$. Then, calling $J_{in} = (u_d+\mu_{in}) f(\mu_{in})$ 
the entering flux, the outgoing one will be given by
\begin{equation}
\label{first}
J_{out}(\mu_{out}) = \int_{-u_d}^1 d\!\mu_{in} P_d(\mu_{in}, \mu_{out})
J_{in}(\mu_{in}) + G(p, \mu_{out}) \equiv P_d \star J_{in}+G_{out}
\end{equation}
where we have also included in the outgoing flux the part deriving from
the injection $G_{out} = G_{inj}(\mu_{out})$ for $\mu < -u_d$.
The symbolic notation will be helpful in the following.

Before specifying the equivalent relationship for the upstream part of the
fluid, we pause to establish the connection between $P_d$ and the solutions
of Eq. \ref{refer} (or Eq. \ref{refer2}). Given the definition of $P_d$,
it seems obvious enough that $P_d$ is the particle flux $(u+\mu) f(\mu)$
given by the solution of Eq. \ref{refer} (or of Eq. \ref{refer2}) with boundary
condition $(u_d+\mu)f = \delta(\mu-\mu_{in})$ (unit entering flux) for $\mu
> -u_d$ at the shock ($z = 0$), and $f$ regular at downstream infinity. In
fact, we can see  that this solution represents a monochromatic (in energy)
flux of particles, all moving initially in the same direction, subject to pitch
angle scattering, which is in turn responsible for kicking them back toward the
shock, occasionally, along a direction $\mu_{out}$, with nonuniform
probability.

One way of looking at $P_d$ is by realizing that it is somewhat related to the
Green's function for a diffusive problem. As an analogy, consider the
paradigmatic diffusion equation
\begin{equation}
\label{heat}
\frac{\partial n}{\partial z} = \frac{\partial^2 n}{\partial \mu^2}
\end{equation}
which is solved by convolving the given boundary conditions with its
Green's function
\begin{equation}
2(\pi z)^{-1/2}\exp(-(\mu-\mu_i)^2/4 z) \rightarrow \delta(\mu-\mu_i)
\end{equation}
as $z\rightarrow 0$. Here too, the Dirac's delta appears in the {\bf boundary
conditions}, and is smoothed out as the time variable ($z$) evolves. The major
difference is that suitable boundary conditions for the above equation
include the whole range in $\mu$, at $z = 0$, while in our problem we can
specify boundary conditions only for $\mu > -u_d$. The function $P_d$ is
the projection of the solution of the problem on exactly that part of the
boundary at $z =0$ where we cannot specify the boundary conditions, a part
which does not exist for Eq. \ref{heat}. Still, because of this relationship,
we will refer to $P_d$ and to $P_u$, shortly to be determined, as the Green's
functions of the problem.

As an aside, we show that, under some physically realistic conditions,
$P_d$ does not depend upon the particle impulse
$p$. This may be at first surprising because both coefficients $d(\mu)$ and
$w(\mu,\mu')$ in Eq. \ref{refer} are allowed to depend on $p$: the proof follows.
Rewrite Eq. \ref{refer} outside the shock (which
means that we can drop the injection term) without the term depending upon the
magnetic field, using as a new variable $y \equiv d(\mu) z$; this can always be
done because, as noticed above, $d>0$. We obtain
\begin{equation}
\label{aux}
\gamma (u+\mu) \frac{\partial f}{\partial y} = - f + \int g(\mu,\mu') f(\mu')
d\!\mu'\:.
\end{equation}
Since we assumed that the $p$--dependence in $w$ and $d$ factors out,
we see from Eq. \ref{constraint} that $g$ is independent of $p$.
Then from Eq. \ref{aux} we see that
$p$ has disappeared from the problem altogether, and that any solution
can depend on $p$ only through the term $y = d(\mu) z$. However, $P_d$ is
the flux at $z = 0$, the shock position, so that we see that when this is
computed ($y = 0$), all dependence on $p$ drops out: $P_d$ does not depend
on $p$. This result was known to Bell (1978), who proved it for Newtonian
shocks; the proof given here is however valid for all shock speeds.

Upstream, we shall also have a probability distribution
$P_u(\mu_{out},\mu_{in})$ that a particle leaving downstream along a direction
$\mu_{out}$ will recross the shock along a direction $\mu_{in}$.
An argument entirely identical to the one given above shows that, often,
$P_u$ does not depend upon the particle's impulse. There is however a
small complication, in that particles of impulse $p_i$ in the downstream frame,
exiting it along a direction $\mu_{out}$ and re-entering it along a
direction $\mu_{in}$, emerge with a different impulse, $p$ (as measured again
in the downstream frame) given by
\begin{equation}
\label{ampl}
p = \frac{1-u_r \mu_{out}}{1-u_r \mu_{in}} p_i \equiv G p_i\;.
\end{equation}
$G$ is the energy amplification a particle receives, as it cycles once around
the shock. Since $-1\leq \mu_{out}\leq -u_d$, and $-u_d\leq \mu_{in}\leq 1$, we
have
\begin{equation}
1 \leq G \leq \frac{1+u_r}{1-u_r}\;.
\end{equation}
Thus, particles leaving the downstream section with impulse $p_i$
provide the reentering flux at a different impulse. A trivial
computation yields the entering flux at impulse $p$ as
\begin{eqnarray}
\label{second}
J_{in}(p, \mu_{in}) = \int_{-1}^{-u_d} d\!\mu_{out} P_u(\mu_{out}, \mu_{in})
\left(\frac{1-u_r \mu_{in}}{1-u_r \mu_{out}}\right)^3
J_{out}(p_i, \mu_{out})
\nonumber \\
+G_{in}(p,\mu_{in}) \equiv P_u\star J_{out}'+G_{in}(p,\mu_{in})
\end{eqnarray}
where again we added to the incoming flux the injected one
$G_{in} = G_{inj}(\mu_{in})$ for $\mu_{in}>-u_d$, and the prime over
$J_{out}$ reminds us that the outgoing flux is to be computed at particle energy
$p_i$ given by Eq. \ref{ampl}.

A comment on normalization is in order. All particles that enter
upstream will in due time be brought backwards across the shock
by the fluid's advection, so that we expect
\begin{equation}
\label{norm}
\int_{-u_d}^1 d\!\mu_{in} P_u(\mu_{out}, \mu_{in}) = 1
\end{equation}
while the analogous statement for the downstream Green's function $P_d$
does not hold for the same reason: a fraction of all particles will be
advected toward downstream infinity, and the returning flux will be smaller
than the departing one. In general, thus, $\int_{-1}^{-u_d} d\!\mu_{out}
P_d(\mu_{in}, \mu_{out}) < 1$. The average probability of coming back to
cross the shock is given by averaging $P_d$ over the whole incoming flux:
\begin{equation}
P = \frac{\int_{-1}^{-u_d} d\!\mu_{out} \int_{-u_d}^1 d\!\mu_{in}
P_d(\mu_{in},\mu_{out}) J_{in} (\mu_{in})}{
\int_{-u_d}^1 d\!\mu_{in} J_{in}(\mu_{in})}\;.
\end{equation}
When the contribution of injection can be neglected, we see from
the above, and from Eq. \ref{first} that
\begin{equation}
\label{prob}
P = \frac{\int_{-1}^{-u_d} d\!\mu_{out} J_{out} (\mu_{out})}{
\int_{-u_d}^1 d\!\mu_{in} J_{in}(\mu_{in})}\;.
\end{equation}
The reason for this is quite simple: in a steady state, like that assumed here,
there can be no accumulation of particles except at infinity; since there
is a larger flux entering the downstream region than leaving it, there must be
an accumulation of particles in it, which can thus be realized only at
downstream infinity, \ie, by leaving the region of the shock.

Our cardinal equations, Eqs. \ref{first},\ref{second}, can now be simply
combined to obtain
\begin{equation}
\label{jin}
J_{in} = P_u\star P_d \star J_{in}' + (P_u\star G_{out}' + G_{in})
\equiv Q\star J_{in}' + X\;.
\end{equation}
This equation can be solved iteratively, using the term in parentheses as the
first guess:
\begin{equation}
J_{in,(0)} = X \;; J_{in,(1)} = Q\star X' + X\;,
J_{in,(N)} = Q\star J_{in,(N-1)}' + X \;...
\end{equation}
to arrive at the solution in the form
\begin{equation}
\label{sol}
J_{in} = X + Q \star X' + Q\star Q\star X'' + Q\star Q\star Q\star X''' +....
\end{equation}
The interpretation of this equation is simple:
$X$ is the flux of injected particles, including both those that are
directly moving in and those that were initially moving out, but have
been turned back exactly once, The operator $Q$ transforms an ingoing flux
into another
ingoing flux, made of the fraction of all particles that have completed
exactly one loop around the shock. Thus the terms $Q\star Q\star X''$ and
so on represent the particles which have made $2,3,...N,...$ loops
around the shock, after injection. The fact that this theory manages to
recover the well--known fact that the spectrum is the superposition of
infinite bumps, each corresponding to particles that have looped an integer
number of times around the shock, can be considered a useful test which has 
been successfully cleared. It also has the advantage of yielding
the distribution function at the shock for the first time both very close to
the injection energy, and very far from it. Also, it substantiates physical
intuition that the solution of the problem must depend only on the injected
flux ($X$) and on the scattering properties of the medium. Lastly, the
equation concerning $J_{out}$ carries no new information: once $J_{in}$
has been found, it can be plugged into Eq. \ref{first} which yields a
$J_{out}$ which automatically satisfies Eq. \ref{second}.

\section{The large-$p$ limit}{}

We see from
Eq. \ref{ampl} that after half a cycle (only the upstream section
changes the particles' energies), the energy of a particle just injected
with energy $p_{inj}$ may span the range ($-1\leq\\mu_{out}\leq -u_d\leq
\mu_{in}\leq 1$)
\begin{equation}
p_{inj} < p < p_{inj} \frac{1+u_r}{1-u_r} \equiv p_{max}\;.
\end{equation}
This shows that the evolution of that part of the particle distribution
function which describes particles making many loops around the shock, must
extend toward larger and larger energies; in fact, since the minimum
amplification is $G = 1$, the process under discussion does not describe
a spreading of the injection distribution around the initial value,
but a drift towards larger and larger energies (plus a spreading, of
course): Fermi acceleration of Type I. So the question naturally arises of
what is the asymptotic spectrum of particles, as we consider asymptotically
larger energies.

If injection provides particles of greatest energy $p_{inj}$, we may neglect
the injection terms in the above equations for energies exceeding $p_{max}$,
and we thus consider Eq. \ref{jin} in this case:
\begin{equation}
\label{noinj}{}
J_{in}(p,\mu) =
\int_{-1}^{-u_d} d\!\mu_{out} P_u(\mu_{out}, \mu)
\left(\frac{1-u_r\mu}{1-u_r\mu_{out}} \right)^3
\int_{-u_d}^1 d\!\xi P_d(\xi, \mu_{out})
J_{in}( \frac{1-u_r\mu_{in}}{1-u_r\mu_{out}}p, \xi)\;.
\end{equation}

We show now that this equation, except for the null solution, has a
solution which is a simple
power law in $p$. Let us first consider two situations, where injection
occurs at slightly different energies, $p_0$ and $p_0 +\delta\!p_0$. At
large energies, $p> p_{max}$, the spectra of these two situations
will differ by very little, $\partial J_{in}/\partial p_0\; \delta\!p_0$.
The difference between two solutions of the inhomogeneous problem ({\it
i.e.}, including injection) must be a solution of the associated
homogeneous problem (the one without injection, Eq. \ref{noinj}), so
that $\partial J_{in}/\partial p_0$ is a solution of Eq. \ref{noinj}.
For obvious dimensional reasons, $J_{in} = h(p/p_0,\mu)/p^3$ so that
(using $\dot{h}(z,\mu) \equiv \partial h/\partial z$)
\begin{equation}
\frac{\partial J_{in}}{\partial p} = -\frac{3 h}{p^4} + \frac{\dot{h}}{p^3 p_0}
\end{equation}
\begin{equation}
\frac{\partial J_{in}}{\partial p_0} = - \frac{\dot{h}}{p^2 p_0^2}
\end{equation}
and, upon eliminating $h$,
\begin{equation}
p \frac{\partial J_{in}}{\partial p} = - 3 J_{in} - \frac{1}{p_0}
\frac{\partial J_{in}}{\partial p_0}
\end{equation}
which shows
$p\partial J_{in}/\partial p$
to be the linear combination
of two solutions of the homogeneous equation \ref{noinj} (for $p>p_{max}$),
and thus a solution itself. If we now assume Eq. \ref{noinj} to have
a unique solution, we find that $p\partial J_{in}/\partial p$ must
be proportional to $J_{in}$, when injection can be neglected. We thus have
\begin{equation}
\frac{1}{J_{in}(p,\mu)} \frac{\partial J_{in}}{\partial p}(p,\mu) =
- \frac{s}{p}
\end{equation}
where $s$ is a constant yet to be determined, and the obvious solution
\begin{equation}
\label{proved}
J_{in}(p,\mu) = \frac{(u_d+\mu) g(\mu)}{p^s}
\end{equation}
follows, where we have arbitrarily factored out $u_d+\mu$ to remind us that
$J$ is a particle flux; thus, $g(\mu)/p^s$ is the particle distribution
function. It must be remarked that this power--law dependence has been
derived without assuming that $P_d$ and $P_u$ are independent of the
particles' momentum $p$.

The importance of the above is obvious: it shows that the particle distribution
function is a power law in the particles' momentum, for every shock speed. This
result had been obtained before in all numerical solutions for arbitrary
shock speeds, and in analytic form in the Newtonian limit, but in the full
relativistic regime it was put in by hand (Kirk and Schneider 1987), and the
boundary conditions could have equally well been satisfied by any other function
with a free paramemter. The present derivation is universal, and it has
the rather unexpected result (at least, for this author) that the power
law solution is exact, not asymptotic, as soon as we consider energies exceeding
the injection ones, \ie, for $p > p_{max}$, and not just $p\gg p_{max}$.
{}

With  the result provided by Eq. \ref{proved}, our principal equations, Eqs.
\ref{first} and \ref{second}, assume the following form, when injection may
be neglected:
\begin{equation}
\label{firsttransf}
(u_d + \mu) g(\mu) = \int_{-u_d}^1 d\!\xi P_d(\xi,\mu) (u_d+\xi) g(\xi)
\end{equation}
\begin{equation}
\label{secondtransf}
(u_d+\mu) g(\mu) = \int_{-1}^{-u_d} d\!\mu_{out} P_u(\mu_{out}, \mu)
\left(\frac{1-u_r\mu}{1-u_r\mu_{out}} \right)^{3-s}
(u_d+\mu_{out}) g(\mu_{out})\;.
\end{equation}
The first one concerns the outgoing flux (with respect to the downstream
section), and thus is valid for $-1\leq\mu\leq
-u_d$, while the second one concerns the entering flux, and applies for
$-u_d\leq\mu\leq 1$. Also, use of Eq. \ref{proved} in Eq. \ref{prob} shows
that the probability of reaching downstream infinity does not depend
upon the particle's momentum, $p$.

Again, they can be combined to obtain
\begin{equation}
\label{fredholm}
(u_d+\mu) g(\mu) = \int_{-u_d}^1 d\!\xi Q^T(\xi,\mu) (u_d+\xi) g(\xi)
\end{equation}
where
\begin{equation}
Q^T(\xi,\mu) \equiv
\int_{-1}^{-u_d} d\!\nu P_u(\nu, \mu) P_d(\xi,\nu)
\left( \frac{1-u_r\mu}{1-u_r\nu}\right)^{3-s}\;.
\end{equation}
which applies to the incoming flux only, $-u_d\leq \mu\leq 1$. Once again, the
outgoing flux carries no new information: if the above is solved, plugging
the solution into Eq. \ref{firsttransf} returns an outgoing flux which
automatically solves Eq. \ref{secondtransf}. All of the problem's information
is in one of the subintervals only, even in the case of the homogeneous
equation.

The above equation, for an arbitrary value of $s$, does {\bf not} have a solution.
The equation which {\bf does} have a solution is
\begin{equation}
\label{truefredholm}
\lambda (u_d+\mu) g(\mu) = \int_{-u_d}^1 d\!\xi Q^T(\xi,\mu) (u_d+\xi) g(\xi)\;.
\end{equation}
In fact, the above equation is an homogeneous equation of the Fredholm
type, with smooth and bounded kernel over a compact interval (Courant and Hilbert
1953), which has at most a finite number of solutions, each belonging to
an eigenvalue $\lambda$ which is to be determined simultaneously with the
eigenvector $g(\mu)$. We thus see that what fixes $s$ is the requirement
that $\lambda = 1$:
the physical value of $s$ is that which has a unitary eigenvalue. We are forced
by our limited mathematical skills to assume that Eq. \ref{fredholm} has
exactly one solution, neither more nor fewer, but, apart from this, the task
of determining the particle distribution function is completed: Eq. \ref{truefredholm}
with $\lambda = 1$ yields the particle distribution function at the shock,
and the all--important particle spectral index $s$, even in the case of
arbitrary particle anisotropy.

\subsection{The Newtonian limit}

In order to gain physical insight into the condition that fixes $s$ ($\lambda = 1$),
we rewrite Eq. \ref{truefredholm} somewhat. First, we write Eq. \ref{secondtransf} as
\begin{equation}
\label{pastrugnata}
\lambda (u_d+\mu) g(\mu) = \int_{-1}^{-u_d} d\!\mu_{out} P_u(\mu_{out}, \mu)
\left(\frac{1-u_r\mu}{1-u_r\mu_{out}} \right)^{3-s}
(u_d+\mu_{out}) g(\mu_{out})\;.
\end{equation}
This can now be integrated over the whole range in $\mu$, $ -u_d\leq \mu\leq 1$,
and divided by the whole outgoing flux, to obtain
\begin{equation}
\label{auxxx}
\lambda \frac{\int_{-u_d}^1 d\!\mu (u_d+\mu)g(\mu)}{\int_{-1}^{-u_d} d\!\mu_{out}
(u_d+\mu_{out}) g(\mu_{out})} =
\frac{\int_{-u_d}^1 d\!\mu
\int_{-1}^{-u_d} d\!\mu_{out} P_u(\mu_{out}, \mu)
\left(\frac{1-u_r\mu}{1-u_r\mu_{out}} \right)^{3-s}(u_d+\mu_{out}) g(\mu_{out})
}{\int_{-1}^{-u_d} d\!\mu_{out} (u_d+\mu_{out}) g(\mu_{out})}
\end{equation}
The term on the left hand side can be rewritten by means of Eq. \ref{prob}
as $\lambda /P$. The term on the right hand side is clearly the average
value of the $(s-3)$--th power of the amplification $G$ (Eq. \ref{ampl}), over the
whole particle distribution: $<G^{s-3}>$. The above is thus
\begin{equation}
\lambda = P <G^{s-3}>\;,
\end{equation}
and demanding that $\lambda = 1$ implies that
\begin{equation}
\label{aux2}
P <G^{s-3}> = 1\;.
\end{equation}
In the Newtonian limit, in which $G-1 \ll 1$, we obviously have $<G^{s-3}>
\approx <G>^{s-3}$, and thus the energy spectral index $k = s-2$ is given by
\begin{equation}
\label{bell}
s-2 \equiv k = 1- \frac{\log P}{\log <G>}
\end{equation}
which is exactly Bell's equation for the spectral index. We have thus established
that the condition $\lambda = 1$ is equivalent to $P <G^{s-3}> = 1$, which
appears to be the correct relativistic generalization of Bell's Newtonian results.

In the Newtonian limit, where $g(\mu) \approx$ constant, independent of $\mu$
(the isotropic limit), it is easy to derive from Eq. \ref{prob} that
$P \approx 1-u$, and from Eq. \ref{auxxx} that $<G> \approx 1+u$, from
which $s = 4$, as per Bell's results for a strong shock.

There are some differences between the fully relativistic approach and
the Newtonian one. First, in the relativistic approach in general
$<G^{s-3}>$ cannot be approximated by $<G>^{s-3}$, contrary to
Peacock's (1981) arguments. Second, here $G$ and $P$ are functions
of $s$, so that Eq. \ref{aux2} sets up a transcendental
equation for $s$. Third, Eq. \ref{aux2} is only of symbolic value,
since it requires knowledge of the anisotropic particle distribution
function, which is not known {\it a priori}. But luckily,
the full relativistic approach developed
here returns simultaneously the anisotropic distribution function (Eq.
\ref{fredholm}).

{}
\subsection{Bits and pieces}

A useful identity is obtained by rewriting Eq. \ref{second} without the
injection term as
\begin{equation}
\frac{J_{in}(p, \mu_{in})}{(1-u_r\mu_{in})^3} =
\int_{-1}^{-u_d} d\!\mu_{out} P_u(\mu_{out}, \mu_{in})
\frac{J_{out}(p_i, \mu_{out})}
{(1-u_r \mu_{out})^3}\;,
\end{equation}
integrating over
$\mu_{in}$, using Eq. \ref{norm}, and then inserting Eq. \ref{proved},
to obtain
\begin{equation}
\label{identity1}
\int_{-u_d}^1 d\!\mu_{in} (1-u_r \mu_{in})^{s-3} (u_d+\mu_{in}) g(\mu_{in}) =
\int_{-1}^{-u_d} d\!\mu_{out} (1-u_r \mu_{out})^{s-3} (u_d+\mu_{out})
g(\mu_{out})
\end{equation}
which of course is not trivial, because the intervals of integration are
different.

It is worth noticing that, in all of Section 3,
no reference whatsoever has been made to the explicit
form of the right hand side of Eq. \ref{refer}; all that is necessary is to
determine the equation's Green's functions $P_d$ and $P_u$, but otherwise all
results apply to all right hand side members, and can then be seen as a
property of the left hand side member $(u+\mu)\partial f/\partial z$. They
are thus of very general validity.

Lastly, we have assumed that Eqs. \ref{fredholm} and \ref{truefredholm}
has one, and just one solution.
If it had none, then the steady state problem would have no solution, and
the problem would be intrinsically time--dependent. If it had several, then
a rather interesting case would arise, whereby shocks moving at the same speed
in the same
media, but with different injection properties would have different
asymptotic spectral indices $s$. The near universality of the index $s$
points of course in the opposite direction, but it is also true that a proof
of the existence of a unique solution is beyond our modest abilities.

\section{$P_u$ and $P_d$}

In this subsection we show how to build $P_u$ and $P_d$. For sake
of definiteness, we concentrate on $P_u$, but a wholly analogous
treatment holds for $P_d$.

We now wish to find an explicit expression for $P_u(\mu_{out},\mu_{in})$.
That this quantity exists, it is {\bf physically} obvious: it represents the
probability that a net flux of particles leaving the downstream section
along a direction $\mu_{out}$ (in downstream variables), reenters along a
direction $\mu_{in}$. We can imagine an experiment to determine this
quantity: an experimenter located next to the shock with a small
cannon shooting out particles along the direction $\mu_{out}$, may
collect them on a screen, determining how many come back along $\mu_{in}$.
It corresponds to a solution of Eq. \ref{refer} or \ref{refer2}
without the injection term, with boundary condition
\begin{equation}
\label{initcon}
d\!J = \delta(\mu_d - \mu_{out}) d\!\mu_d
= \delta(\mu_u - \mu^{(u)}_{out}) d\!\mu_u
\;.
\end{equation}
Here, $\mu^{(u)}_{out}$ is the value of $\mu_{out}$ in the upstream
frame variables, and the last identity is a trivial property of
Dirac's deltas.

In order to find $P_u$, we now use separation of variables to solve
Eq. \ref{refer} or Eq. \ref{refer2}. For instance, writing
$f(\mu_u,z) = A(z) B(\mu_u)$, we find the solutions of
\begin{equation}
\gamma \frac{1}{A(z)}\frac{d A}{dz} = \lambda_n =
\frac{1}{(u+\mu_u) B(\mu_u)}
\frac{d}{d \mu_u}\left(D(\mu_u,p_u) (1-\mu_u^2)
\frac{d B}{d\mu_u}\right)\:.
\end{equation}
Calling $B_n$ the angular eigenvectors, we find a generic solution
of the above as
\begin{equation}
\label{test}
f = \sum_n a_n \exp(\lambda_n z /\gamma) B_n(\mu_u)
\end{equation}
where the $a_n$'s are coefficients which we must choose so
as to satisfy the initial conditions, Eq. \ref{initcon}.
We must thus have, at $z =0$,
\begin{equation}
\gamma (u+\mu_u) f = \sum_n a_n(\mu_{out}^{(u)})
(u+\mu_u) \gamma B_n(\mu_u)
= \delta(\mu_u-\mu^{(u)}_{out})\;.
\end{equation}
Obviously, the coefficients $a_n$'s will depend upon the
point where the delta is located, $\mu_{out}^{(u)}$.

Here there is a tricky, but important point. The above eigenvector
problem is well--known to have solutions belonging to both positive
and negative eigenvalues (plus the $\lambda = 0$ case). Eigenvectors
belonging to negative values of $\lambda_n$, of which there is
an infinite number, are physically ill--behaved at upstream infinity,
as can be seen from Eq. \ref{test}, so that we surely have to
restrict ourselves, in all sums above to the well--behaved
eigenvectors, {\it i.e.}, those with $\lambda_n > 0$. But does
the above equation then hold? The answer is yes, as we now show.

The fact that an
infinite set of functions can be arranged to satisfy the
above equation is called by mathematicians {\bf completeness},
indicating that any function can be written as the superposition
of $B_n$'s with suitable coefficients. If the above equation
holds, in fact, multiplying both sides by any function
$F(\mu_{out}^{(u)})$ and integrating over the whole range of
$\mu_{out}^{(u)}$, we find that
\begin{equation}
F(\mu_u) = \sum_n \left( \int d\!\mu_{out}^{(u)} a_n(\mu_{out}^{(u)})
F(\mu_{out}^{(u)})\right) \gamma (u+\mu_u) B_n(\mu_u)\;,
\end{equation}
proving that we can write {\bf any} function as the superposition
of functions from the set, with suitable coefficients, and thus that
the set is, by definition, complete.
Now, the set of all eigenfunctions $B_n$ is
well--known to be complete, but here is the rub: the full set includes
both the eigenvectors which have $\lambda >0$ (which thus
are physically well behaved at upstream infinity) {\bf and} those
which have $\lambda \leq 0$, which diverge at upstream infinity, and
which we thus cannot use.

The half set of the well--behaved eigenvectors is not complete over
the whole range in $\mu$, but
here we are helped by an unusual property of the eigenvectors of this
equation: it can be shown in fact (Freiling, Yurko and Vietri 2002) that the
half--set of all well--behaved eigenvectors (i.e., those with $\lambda
> 0$) is complete in the restricted range $-1\leq \mu_u \leq -u$.
This property (sometimes called half-range half-completeness) implies
that, for $\mu_u\leq -u$, we can always write
\begin{equation}
\gamma (u+\mu_u) f = \sum'_n a_n (u+\mu_u) \gamma B_n(\mu_u)
= \delta(\mu_u-\mu^{(u)}_{out})
\end{equation}
where however now the prime over the sum reminds us that the summation
is extended only to well--behaved eigenvectors. For $\mu_u>-u$,
the above equation is {\bf not} verified, and the sum will give
some function, which we can now identify with our required $P_u$.
Thus, $P_u$ is the continuation to the rest of the interval ($\mu_u
> -u$) of the sum above, which inside the restricted range $\mu \leq -u$
satisifes the above equality.

One word of caution should be stated here: the property of half-range
half-completeness has been proved for the eigenvectors of Eq. \ref{refer2},
but not for those of Eq. \ref{refer}. Still, since they are known
to be so similar (both are called of the third, or polar type, one
within the realm of integral equations, the other one of differential
equations, to emphasize several formal similarities), and since the
above physical argument makes it extremely plausible that $P_u$ exists
also for this case, we shall make no further distinction between the
two cases, and proceed as if this argument went through for both
Eq. \ref{refer2} and Eq. \ref{refer}.

We now find the coefficients $a_n$'s. The well--behaved eigenvectors
are not orthormal over the restricted range $\mu_u \leq -u$. They
can easily made so by means of a well--known procedure (Schmidt's
diagonalization, Courant and Hilbert 1953), which leaves the lowest
order eigenvector unaffected (except for the normalization), and suitably
modifies the others in a finite number of steps. Once this
new basis $P_n$ has been found, since it is complete (from the
half-range half completeness), we easily find, for $\mu_u\leq -u$,
\begin{equation}
(u+\mu_u) \sum_n P_n(\mu^{(u)}_{out}) P_n(\mu_u)
= \delta(\mu_u-\mu^{(u)}_{out})
\end{equation}
and thus, by working backwards to Eq. \ref{fluxdef},
(for $\mu_{out} < -u_d$, $\mu_{in} > -u_d$)
\begin{equation}
\label{expansion}
P_u(\mu_{out},\mu_{in}) d\!\mu_{in} =
(u+\mu_u) \sum_n P_n(\mu^{(u)}_{out}) P_n(\mu_u) d\!\mu_u\;.
\end{equation}

\section{Summary}

The major results presented in this paper are the following:
\begin{itemize}
\item
We have presented a relativistically covariant equation (Eq. \ref{refer})
for the particle distribution function, which includes pitch--angle
scattering (without assuming that the diffusive approximation holds), the
effect of smooth magnetic fields, and particle injection at the shock;
this equation applies separately in the upstream and downstream frames.
\item
We have described the transport properties of the upstream and
downstream media by means of two Green's functions, $P_d$ and $P_u$,
which are independent of the particle distribution functions, and
which have been explicitly constructed in terms of the eigenfunctions
of the angular part of the problem (Eq. \ref{expansion}).
\item
We have established two relations (Eq. \ref{first} and \ref{second})
which provide the true boundary conditions of the problem, giving
the flux entering the downstream section in terms of the one
leaving the upstream section, and {\it vice cersa}, plus the
injection terms. This naturally sets up an equation for the
particle flux, Eq. \ref{jin}.
\item
We have solved this equation, Eq. \ref{sol}, in a general way, which yields
the particle spectrum both close to and far from the injection energies.
\item
We have shown that under physically realistic conditions, the probabilities
of returning to the shock, $P_u$ and $P_d$, do not depend upon the particles'
energy $p$.
\item
In the limit of large particle energies, we have shown that the
spectrum is a pure power law in the particles' momentum, even when
the probabilities of returning at the shock, $P_u$ and $P_d$,
depend on $p$ (Eq. \ref{proved}).
\item
In this limit, we have then simplified our main equations, arriving at
a system of equations (Eq. \ref{firsttransf}, \ref{secondtransf}, or their
combination, Eq. \ref{fredholm}), and further showing that the requirement
that Eq. \ref{fredholm} has the eigenvalue $\lambda = 1$ fixes the
all--important asymptotic particle spectral index $s$.
\item
We have recovered the low speed, Newtonian limit of Bell (1978) and
Blandford and Ostriker (1978), and provided a relativistic generalization
(Eq. \ref{aux2}) which is however of little use because it requires
knowledge of the particle anisotropic distribution function, which can
be obtained only through the full solution of Eq. \ref{fredholm}, which
automatically incorporates Eq. \ref{aux2}.
\end{itemize}

Though the treatment presented in this apper may appear rather abstract,
it is not without practical consequences. For instance, the fact that
the asymptotic spectral index $s$ is not asymptotic, but it applies
as soon as $p > p_{max} = p_{inj} (1+u_r)/(1-u_r)$, where $u_r$ is the modulus of
the relative speed between the upstream and downstream fluids, implies
that numerical simulations trying to compute $s$ need not extend to very
large particle energies, but can save precious computational time by sticking
to energies exceeding $p_{max}$ by a small factor, dictated exclusively by
numerical questions. Also, the treatment presented in Section 3 may be somewhat
simplified by a judicious use of Fredholm's formulae, resulting in a considerable
simplification for the calculational side of the theory. And lastly, they allow 
the treatment of the hyperrelativistic limit, as will be shown in a future paper.

{}

\end{document}